\documentclass[preprint,aps,floatfix,showpacs]{revtex4}
\usepackage{dcolumn}
\usepackage{bm}
\usepackage{graphicx}
\begin{document}
\title{Variation of fundamental constants in space and time: theory and observations}
\author{V. V. Flambaum$^{1,2}$}
\affiliation{
$^1$\mbox{ Institute for Advanced Study,
Massey University (Albany Campus),Private Bag 102904, North Shore MSC
 Auckland, New Zealand} \\
$^2$\mbox{School of Physics, University of New South Wales, Sydney 2052,
           Australia}}
\begin{abstract}
Review of recent works devoted to the temporal and spatial
variation of the fundamental constants and dependence of the fundamental
constants on the gravitational potential (violation of local position invariance) is presented.
 We discuss the variation of the fine structure constant $\alpha=e^2/\hbar c$, strong interaction and
fundamental masses (Higgs vacuum), e.g. the electron-to-proton mass ratio
$\mu=m_e/M_p$ or  $X_e=m_e/\Lambda_{QCD}$ and  $X_q=m_q/\Lambda_{QCD}$.
We also present new results from Big Bang nucleosynthesis
and Oklo natural nuclear reactor data 
and propose new measurements of enhanced effects in atoms, nuclei and molecules,
 both in quasar and laboratory spectra.
\end{abstract}
\maketitle
\section{Introduction}
\label{intro}
Theories unifying gravity with other interactions suggest temporal and
spatial variation of the fundamental ``constants'' in expanding Universe 
(see e.g. review \cite{uzan}).
The spatial variation can explain  fine tuning of the fundamental
constants which allows humans (and any life) to appear. We appeared in
the area of the Universe where the values  of the fundamental constants
are consistent with our existence. The fundamental constants may be slightly
different near massive bodies (see e.g. review \cite{shuryak}).
 There are some hints for
 the variation of  different fundamental constants in  quasar
absorption spectra \cite{webb1,webb2,murphy1,murphy2,murphy3,reynold}  and Big Bang nucleosynthesis \cite{dmitriev,wiringa} data.
However, a majority of publications report limits on the variations (see e.g. recent reviews \cite{flambaumint,lea}).

 The hypothetical unification of all interactions implies that  variations of different
fundamental constants may be related \cite{marciano,calmet,langacker,wet,dent}.
We can only detect variation of dimensionless fundamental constants.
We will discuss variation of the fine structure constant  $\alpha$ and dimensionless
ratios  $X_e=m_e/\Lambda_{QCD}$ and  $X_q=m_q/\Lambda_{QCD}$ where $m_e$ and $m_q$ are the electron
and quark masses, and  $\Lambda_{QCD}$ is  the quantum chromodynamics
(QCD) scale  (defined as the position of the Landau pole in
the logarithm for the running strong coupling constant,
$\alpha_s(r) \sim 1/\ln{(\Lambda_{QCD} r/\hbar c)}$). The proton mass $m_p$ is proportional
to $\Lambda_{QCD}$, therefore the relative variation of $\mu=m_e/M_p$ is equal to the relative
variation of  $X_e=m_e/\Lambda_{QCD}$ (if we neglect a small contribution of quark masses ($m_q\sim 5$ MeV)
to the proton mass, $m_p=938$ MeV). In the Standard model electron and quark masses are proportional
to the vacuum expectation value of the Higgs field.
 
A simple estimate of the relations between the variations of different fundamental constants may be
obtained using the idea of Grand Unification.
The strong (i=3), and electroweak (i=1,2) inverse coupling
constants have the following
dependence on the scale $\nu$ and normalization point $\nu_0$:
\begin{equation} \label{eqn_inv_alpha}
 \alpha_i^{-1}(\nu)=\alpha_i^{-1}(\nu_0)+b_i ln(\nu/\nu_0)
\end{equation}
In the Standard Model, $2\pi b_i=41/10,-19/6,-7$;
the electromagnetic $\alpha^{-1}=(5/3)\alpha_1^{-1}+\alpha_2^{-1}$
and the strong $\alpha_s=\alpha_3$.
In the Grand Unification Theories (GUT) all coupling constants are equal at the unification scale,
$\alpha_i(\nu_0)\equiv\alpha_{GUT}$.
If we assume that $\alpha_{GUT}$ varies, then Eq.~(\ref{eqn_inv_alpha})
gives us the same shifts for all inverse couplings:
\begin{equation}
 \delta \alpha_1^{-1}=\delta \alpha_2^{-1}= \delta \alpha_3^{-1}=
 \delta \alpha_{GUT}^{-1} \ .
\end{equation}
We see that the variation of the strong interaction constant
$\alpha_3(\nu)$ at low energy $\nu$ is much larger than the variation
of the elecromagnetic constant $\alpha$, since
$\delta \alpha_3/\alpha_3=(\alpha_3/\alpha_{1,2})\delta \alpha_{1,2}/
\alpha_{1,2}$ and $\alpha_3 \gg \alpha_{1,2}$.

The variation of $m/\Lambda_{QCD}$ can be estimated from the definition
of $\Lambda_{QCD}$. The running of $\alpha_s$ near the electroweak scale
is given by
\begin{equation} \label{Lalphas}
 \alpha_s(\nu)^{-1} \approx b_s ln(\nu/\Lambda_{QCD})
\end{equation}
Let us take $\nu=m_z$ where $m_z$ is the $Z$-boson mass. The variation
of eq. (\ref{Lalphas}) and relations above give
\begin{equation}\label{Lalpha}
 \frac{\delta (m_z/\Lambda_{QCD})}{(m_z/\Lambda_{QCD})}=
 -\frac{1}{b_s\alpha_s(m_z)} \frac{\delta\alpha_s(m_z)}{\alpha_s(m_z)}
 \sim \frac{C}{\alpha(m_z)} \frac{\delta\alpha(m_z)}{\alpha(m_z)}
\end{equation}
The value of the constant $C$ here depends on the model used. However,
the enhancement $1/\alpha \sim 100$ should make the factor $C/\alpha$ large.
Note that $m_z$ (as well as $m_{e}$ and $m_{q}$) is proportional to the Higgs vacuum expectation value.

If this  estimate is correct, the variation  in  $X_{e,q}=m_{e,q}/\Lambda_{QCD}$ or $\mu=m_e/M_p$ 
may be easier to detect than the variation in $\alpha$.
The cosmological variation of $m_q/\Lambda_{QCD}$ can be extracted from
the big bang nucleosynthesis (BBN), quasar absorption spectra
and Oklo natural nuclear reactor data~\cite{FlambaumShuryak2002,dmitriev1,FlambaumShuryak2003}.
For example, the factor of three disagreement between the calculations
and measurements of the BBN abundance of $^7$Li may, in principle, be
explained by the variation of $m_q/\Lambda_{QCD}$ at the level of
$\sim 10^{-3}-10^{-2}$ \cite{dmitriev,wiringa} (see also recent work~\cite{BBN}).
The claim of the variation of the fundamental constants based on the
Oklo data in Ref.~\cite{Lam} is not confirmed by recent studies~\cite{Oklo1,Oklo2,Oklo3}
which give a stringent limit on the possible variation of the resonance
in $^{150}$Sm during the last two billion years.
The search for the variation of $m_e/\Lambda_{QCD}$ using the quasar
absorption spectra gave a non-zero result in Ref.~\cite{reynold}
but zero results in Refs.~\cite{tzana,tzana1,FK1}.
The present time variation of $m_{e,q}/\Lambda_{QCD}$
can be extracted from comparison of different atomic~\cite{tedesco}, molecular
or nuclear~\cite{th1,th4} clocks. New enhanced effects have been proposed.

\section{Big Bang Nucleosynthesis}
\label{BBNsec}

The result of our work Ref.~\cite{dmitriev} suggested that a reduced deuteron binding
energy of $\Delta Q/Q = -0.019 \pm 0.005$ would yield a better fit to 
observational data (the WMAP value of barion-to-photon ratio$\eta$ and measured $^2$H, $^4$He, and
$^7$Li abundances) for Big Bang Nucleosynthesis. Using our calculations
\cite{FlambaumShuryak2003} we obtained in Ref.  \cite{dmitriev} an estimate of the strange quark mass variation.

Recently Dent, Stern, and Wetterich~\cite{BBN} calculated the sensitivity of BBN 
abundances for $^{2}$H, $^{4}$He and $^{7}$Li to the variation of binding 
energies of $^{2,3}$H, $^{3,4}$He, $^{6,7}$Li and $^{7}$Be in a linear 
approximation. In the works \cite{wiringa,roberts,roberts1,thomas} we calculated dependence
of these binding energies on the light quark mass variation and
 estimated the sensitivity of  BBN yields to variation of the quark mass.
 Then we used the  observational data to obtain
the following equations for $^{2}$H, $^{4}$He and $^{7}$Li  \cite{wiringa}:
\begin{equation}\label{H}
1 + 7.7 x = \frac{2.8 \pm 0.4}{2.61 \pm 0.04} = 1.07 \pm 0.15 \ ,
\end{equation}
\begin{equation}\label{He}
1 - 0.95 x = \frac{0.249 \pm 0.009}{0.2478 \pm 0.0002} = 1.005 \pm 0.036 \ ,
\end{equation}
\begin{equation}\label{Li}
1 - 50 x = \frac{1.5 \pm 0.5}{4.5 \pm 0.4} = 0.33 \pm 0.11 \ ,
\end{equation}
where $x=\delta X_q/X_q$.  
These equations yield 3 consistent values of $x$: 
$0.009 \pm 0.019$, $-0.005 \pm 0.038$ and $0.013 \pm 0.002$.
The statistically weighted average of $\delta X_q/X_q=0.013 \pm 0.002$ 
is dominated by the $^{7}$Li data.  A more accurate calculation
should take into account the effect of the $^8$Be binding energy variation 
(which is not calculated in Ref.~\cite{BBN}), the variation of the virtual 
$^1$S$_0(np)$ level, and non-linear corrections in $x$ which are important 
for $^{7}$Li.
Allowing for the theoretical uncertanties we should understand this
BBN result as $\delta X_q/X_q=K \cdot (0.013 \pm 0.002)$ where $K \sim 1$,
where the expected accuracy in $K$ is about a factor of 2.  Note that here we
neglected effects of the strange quark mass variation.  A rough estimate
of these effects on BBN due to the deuteron binding energy variation
was made in Refs.~\cite{FlambaumShuryak2003,dmitriev}.

\section{Oklo natural nuclear reactor}
\label{sec:3}

   The results from Oklo natural nuclear reactor are based on the measurement
of the position of very low energy resonance ($E_r=0.1$ eV) in neutron 
capture by $^{149}$Sm nucleus. The estimate of the shift of this resonance
 induced by the
 variation of $\alpha$ have been done long time ago in works \cite{shl,Dyson}.
Recently we performed a rough estimate of the effect of the variation of
  $m_q/\Lambda_{QCD}$ \cite{FlambaumShuryak2002,dmitriev,FlambaumShuryak2003}. The final result is
 \begin{equation}\label{deltaE}
\delta E_r \approx 10^6 eV (
\frac{\delta \alpha}{\alpha} -10 \frac{\delta X_q}{X_q }
+ 100 {\delta X_s \over X_s})
\end{equation} 
where $X_q=m_q/\Lambda_{QCD}$, $X_s=m_s/\Lambda_{QCD}$,
$m_q=(m_u+m_d)/2$ and $m_s$ is the strange quark mass. Refs. \cite{Oklo1,Oklo2,Oklo3}
found that $|\delta E_r| < 0.1$ eV. This gives us a limit 
 \begin{equation}\label{Oklolimit}
|0.01\frac{\delta \alpha}{\alpha} -0.1 \frac{\delta X_q}{X_q }
 +{\delta X_s \over X_s}|<10^{-9}
\end{equation} 
The contribution of the $\alpha$ variation in this equation is very small
and should be neglected since the accuracy of the calculation of the main term
 is low.
Thus, the Oklo data can not give any limit on the variation of $\alpha$.
Assuming linear time dependence during last 2 billion years 
we obtain an estimate
$|\dot{X_s}/X_s| < 10^{-18}$~yr$^{-1}$.

\section{Optical atomic  spectra}
\label{sec:4}

\subsection{Comparison of quasar absorption spectra with laboratory spectra}
To perform measurements of $\alpha$ variation by comparison of cosmic and
 laboratory optical spectra  we developed a new approach
\cite{dzubaPRL,dzuba1999} which improves the sensitivity to a
variation of $\alpha$ by more than an order of magnitude.
  The relative value of any relativistic
corrections to atomic transition frequencies is proportional to
$\alpha^2$. These corrections can exceed the fine structure interval
between the excited levels by an order of magnitude (for example, an
$s$-wave electron does not have the spin-orbit splitting but it has the
maximal relativistic correction to energy). The relativistic corrections
vary very strongly from atom to atom and can have opposite signs in
different transitions (for example, in $s$-$p$ and $d$-$p$
transitions). Thus, any variation of $\alpha$ could be revealed by
comparing different transitions in different atoms in cosmic and laboratory
spectra.

This method provides an order of magnitude precision gain compared to
measurements of the fine structure interval.  Relativistic many-body
calculations are used to reveal the dependence of atomic frequencies on
$\alpha$ for a range of atomic species observed in quasar absorption
spectra \cite{dzuba1999,dzubaPRL,Dy,q1,q2,q3,q4} (a 2004 summary may be found in Ref. \cite{shopping}).
  It is convenient to present results for the
transition frequencies as functions of $\alpha^2$ in the form
\begin{equation}
\label{q1}
\omega = \omega_0 + q  x,
\end{equation}
where $x = (\frac{\alpha}{\alpha_0})^2 - 1 \approx
 \frac{2 \delta \alpha}{\alpha}$ and
 $\omega_0 $ is a laboratory frequency of a particular transition.
We stress that the second  term contributes only if $\alpha$
deviates from the laboratory value $\alpha_0$. 
We performed accurate many-body calculations of the coefficients $q$
for all transtions of astrophysical interest (strong E1 transtions from the
ground state)  in Mg, Mg II, Fe II, Fe I, Cr II, Ni II, Al II, Al III, Si II,
Zn II, Mn II (and many other atoms and ions which have not been used in the quasar measurements yet
because of the absence of accurate laboratory wavelenths - see \cite{shopping}).
 It is very important that this  set of transtions
contains three large classes : positive shifters (large positive coefficients
$q > 1000 $ cm$^{-1}$), negative shifters (large negative coefficients
$q <- 1000 $ cm$^{-1}$) and anchor lines with small values of $q$.
This gives us an excellent control of systematic errors
since systematic effects do not ``know'' about sign and magnitude
of $q$. Comparison of cosmic frequencies $\omega$  and
 laboratory frequencies $\omega_0$ allows us
to measure $\frac{ \delta \alpha}{\alpha}$.

 Three independent samples of data contaning 143
absorption  systems spread over red shift range $0.2 <z < 4.2$.
 The fit of the data gives \cite{murphy1}
 is $\frac{ \delta \alpha}{\alpha}=
(-0.543 \pm 0.116) \times 10^{-5}$. If one assumes the linear dependence
of $\alpha$ on time, the fit of the data gives $d\ln{\alpha}/dt=
(6.40 \pm 1.35) \times 10^{-16}$ per year
 (over time interval about 12 billion years).
 A very extensive search for possible
systematic errors has shown that known systematic effects can not explain
 the result (It is still  not completely excluded that the effect may be
 imitated by a large change of abundances of isotopes
 during last 10 billion years. 
We have checked that different isotopic abundances for any single
element can not imitate the observed effect. It may be an 
improbable ``conspiracy'' of several elements).

 Recently our method and calculations
 \cite{dzuba1999,dzubaPRL,Dy,q1,q2,q3}
were used by two other groups \cite{chand,Levshakov,Levshakov1}. However, they have
not detected any variation of $\alpha$.
  Recently the results of \cite{chand} were questioned in Refs.
 \cite{murphy2,murphy3}. Re-analysis of Ref. \cite{chand}
data revealed flawed parameter estimation methods.
The authors of \cite{murphy2,murphy3} claim that the same spectral data fitted 
more accurately give  $\frac{ \delta \alpha}{\alpha}=
(-0.64 \pm 0.36) \times 10^{-5}$ (instead of  $\frac{ \delta \alpha}{\alpha}=
(-0.06 \pm 0.06) \times 10^{-5}$ in Ref.\cite{chand}). However, even this
revised result may require further revision. 

 Note  that the results of 
\cite{webb1,webb2,murphy1} are based on the data from the
 Keck telescope which is located in the Northen hemisphere  (Hawaii).
 The results of \cite{chand,Levshakov,Levshakov1,murphy2,murphy3}
are based on the data from the different telescope (VLT) located
in the Southern hemisphere (Chile). Therefore, some difference in the results
may appear due to the spatial variation of $\alpha$. 

    Using opportunity I would like to ask for new, more accurate laboratory
 measurements of  UV transition frequencies which have been observed
in the quasar absorption spectra. The ``shopping list'' is presented
in \cite{shopping}. We also need the laboratory measurements of 
isotopic shifts  - see \cite{shopping}. We have performed very complicated
calculations of these isotopic shifts \cite{isotope1,isotope2,isotope3,isotope4,isotope5}. However, the
 accuracy of  these calculations in atoms and ions with open d-shell
(like Fe II, Ni II, Cr II, Mn II, Ti II) may be very low. The measurements
 for at list few lines are needed to test these calculations.
These measurements would be very important for a study of  evolution of
 isotope abundances in the Universe, to exclude the systematic effects
 in the search for 
$\alpha$ variation and to test models of nuclear reactions in stars
and supernovi.

A comparison of the hyperfine transition
in atomic hydrogen with optical transitions in ions, was done in
Refs.~\cite{tzana,tzana1}. This method allows one to study
time-variation of the parameter $F=\alpha^2g_p\mu$, where $g_p$ is
proton $g$-factor. Analysis of 9 quasar spectra with redshifts $0.23
\le z \le 2.35$ gave
\begin{equation}\label{x_var}
    \delta F/F =(6.3\pm 9.9)\times 10^{-6},
\end{equation}
\begin{equation}\label{x_vara}
     \dot{F}/F =(-6\pm 12)\times 10^{-16}~\mathrm{yr}^{-1}.
\end{equation}

\subsection{Optical atomic clocks}
Optical clocks also include transitions which have positive, negative
or small constributions of the relativistic corrections to frequencies.
We used the same methods of the relativistic many-body calculations
to  calculate the dependence on $\alpha$ \cite{dzuba1999,clock1,clock2,clock3,clock4,Dy}.
A 2004 summary of the results for the coefficients $q$ is presented in \cite{clock5}.
 The coefficients
$q$ for optical clock transitions  may be substantially larger than
in cosmic transitions since the clock transitions are often in  heavy atoms
(Hg II, Yb II, Yb III, etc.) while cosmic spectra contain mostly light
atoms lines ($Z <33$). The relativistic effects are proporitional
to $Z^2 \alpha^2$.

\section{Enhanced effects of $\alpha$ variation in atoms}
An  enhancement of the relative effect of $\alpha$ variation can be obtained
in transition between the almost degenerate levels in Dy atom
 \cite{dzuba1999,Dy,clock4}.
These levels move in opposite directions if  $\alpha$ varies. The relative
variation may be presented as $\delta \omega/\omega=K \delta \alpha /\alpha$
 where the coefficient $K$ exceeds $10^8$. Specific
 values of $K=2 q/\omega$ are different for different 
 hyperfine components and isotopes which have different $\omega$;
 $q=30,000$ cm$^{-1}$,  $\omega \sim 10^{-4}$ cm$^{-1}$. 
 An experiment is currently underway to place limits on
$\alpha$ variation using this transition \cite{budker,budker1}.
The current limit is
 $\dot{\alpha}/\alpha=(-2.7 \pm 2.6) \times 10^{-15}$~yr$^{-1}$.
Unfortunately, one of the levels has  quite a large linewidth
and this limits the accuracy.

Several enhanced effects of $\alpha$ variation in atoms have been calculated
in \cite{dzuba5,nevsky}.

\section{Enhanced effect of  variation of $\alpha$
and strong interaction in UV transition of $^{229}$Th nucleus (nuclear clock)}

A very  narrow level  $(7.6\pm 0.5)$ eV above the ground state exists
 in $^{229}$Th nucleus \cite{th7}. The position
of this level was determined from the energy differences of many high-energy
$\gamma$-transitions to the ground and excited
 states. The subtraction  produces
the large  uncertainty in the position of the 7.6 eV excited state.
 The width of this level is estimated to be
about $10^{-4}$ Hz \cite{th2}. This would explain why it is so hard to find
 the direct radiation in this very weak  transition.
 However, the search for the direct radiation continues
\cite{private}. 

  The  $^{229}$Th transition is very narrow and can be investigated
 with laser spectroscopy.
 This makes $^{229}$Th a possible reference for an
 optical clock of very high accuracy, and opens a new possibility
for a laboratory search for the varitation of the fundamental constants
\cite{th4}.

As it is shown in Ref. \cite{th1} there is an additional very important
 advantage.
According to Ref. \cite{th1} the relative effects of variation of
 $\alpha$ and $m_{q}/\Lambda_{QCD}$ are enhanced by 5 orders of magnitude.
This estimate has been confirmed by the recent calculation \cite{He} and preliminary results of  our
new calculations. The accurate results of the calculations will be published soon.
 A rough estimate for the relative variation of the $^{229}$Th
 transition frequency  is
 \begin{equation}\label{deltaf}
\frac{\delta \omega}{\omega} \approx 10^5 (
0.1 \frac{\delta \alpha}{\alpha} +   \frac{\delta X_q}{X_q })
\end{equation} 
where $X_q=m_q/\Lambda_{QCD}$.
Therefore, the Th  experiment would
have the potential of improving the  sensitivity to temporal
variation of the fundamental
constants by many orders of magnitude. 
Indeed, we obtain the following energy shift
in 7.6 eV $^{229}$Th transition:
\begin{equation}\label{delta3}
\delta \omega \approx 
\frac{\delta X_q}{X_q}  MeV
\end{equation}
This corresponds to the frequency shift
$\delta \nu \approx 3\cdot 10^{20} \delta X_q/X_q$ Hz.
The width of this transition is $10^{-4}$ Hz so one may hope
to get the sensitivity to the variation of $X_q$ about $10^{-24}$
per year. This is  $10^{10}$ times better than the current atomic clock
limit on the variation of $X_q$,  $\sim 10^{-14}$ per year.

Note that there are other narrow low-energy levels in nuclei,
 e.g. 76 eV level in $^{235}U$ with the 26.6 minutes lifetime
 (see e.g.\cite{th4}). One may expect a similar  enhancement there.
Unfortunetely, this level can not be reached with usual lasers. In principle,
 it may be investigated using a free-electron laser or synchrotron radiation.
However, the accuracy
of the frequency measurements is much lower in this case.

\section{Atomic microwave clocks}
Hyperfine microwave transitions may be used to search for $\alpha$-variation \cite{prestage}. 
  Karshenboim
 \cite{Karshenboim2000} has pointed out that measurements of ratios
 of hyperfine structure intervals in different atoms are also sensitive to
 variations in nuclear magnetic moments. However, the magnetic moments
are not the fundamental parameters and can not be directly compared with
 any theory of the variations. Atomic and nuclear calculations are needed 
for the interpretation of the measurements. We have performed both
atomic calculations of $\alpha$ dependence \cite{dzuba1999,clock1,clock2,clock3,clock4,clock5,Dy} and
 nuclear calculations of $X_q=m_q/\Lambda_{QCD}$ dependence \cite{tedesco} (see also \cite{thomas})
 for all microwave transitions of current experimental interest including
hyperfine transitions in $^{133}$Cs, $^{87}$Rb, $^{171}$Yb$^+$,
$^{199}$Hg$^+$, $^{111}$Cd, $^{129}$Xe, $^{139}$La, $^{1}$H,  $^{2}$H and
 $^{3}$He. The  results for the dependence of the transition frequencies
 on variation
of $\alpha$, $X_e=m_e/\Lambda_{QCD}$ and  $X_q=m_q/\Lambda_{QCD}$
 are presented in Ref.\cite{tedesco} (see the final results in the Table IV
 of Ref.\cite{tedesco}). Also, one can find there experimental
 limits on these variations  which follow from the recent measurements. 
The accuracy is approaching $10^{-15}$ per year. This may be compared
to the sensitivity $\sim 10^{-5}-10^{-6}$ per $10^{10}$ years obtained using 
the quasar absorption spectra.

    According to Ref. \cite{tedesco} the frequency ratio $Y$ of the 282-nm
 $^{199}$Hg$^+$ optical clock transition to the ground state hyperfine
 transition
in  $^{133}$Cs has the following dependence on the fundamental constants:
\begin{equation}\label{Hg}
\dot{Y}/Y=-6\dot{\alpha}/\alpha -\dot{\mu}/\mu -0.01 \dot{X_q}/X_q
\end{equation}
In the work \cite{clock_1} this ratio has been measured:
$\dot{Y}/Y=(0.37 \pm 0.39) \times 10^{-15}$~yr$^{-1}$.
Assuming linear time dependence we obtained the quasar result \cite{FK1}
$\dot{\mu}/\mu=\dot{X_e}/X_e=(1 \pm 3) \times 10^{-16}$~yr$^{-1}$.
A combination of this result and the atomic clock result \cite{clock_1} for $Y$
 gives the best limt on the variation of $\alpha$:
$\dot{\alpha}/\alpha=(-0.8 \pm 0.8) \times 10^{-16}$~yr$^{-1}$.
Here we neglected the small ($\sim 1\%$) contribution of $X_q$.

\section{Enhancement of variation of
fundamental constants in ultracold atom and molecule systems near
Feshbach resonances}
Scattering length $A$, which can be measured in Bose-Einstein condensate
and Feshbach molecule experiments, is extremely sensitive to the
variation of the
electron-to-proton mass ratio $\mu=m_e/m_p$ or $X_e=m_e/\Lambda_{QCD}$
 \cite{chin}.
\begin{equation}\label{d_a_final}
\frac{\delta A}{A}=K\frac{\delta\mu}{\mu}=K\frac{\delta X_e}{X_e},
\end{equation}
 where $K$ is the  enhancement factor.
For example, for Cs-Cs collisions we obtained
 $K\sim 400$. With the Feshbach resonance, however, one is
given the flexibility to adjust position of the resonance using
external  fields.  Near a narrow magnetic or an optical Feshbach resonance
 the enhancement factor $K$ may be increased by many orders of magnitude. 

\section{Molecular spectra}
Recently we wrote a review about search for the variation of the fundamental
constants in quasar and laboratory molecular spectra \cite{kozlovmol}.
Below I present several examples related to our works.

\subsection{Comparison of hydrogen hyperfine and molecular rotational quasar  spectra}
\label{rot}
The frequency of the hydrogenic hyperfine line is proportional to
$\alpha^2\mu g_p$ atomic units, molecular rotational frequencies are proportional to 
 $\mu$ atomic units. The comparison places limit on the
variation of the parameter $F=\alpha^2 g_p$~\cite{DWB98}. Recently
similar analysis was repeated by Murphy et al \cite{MWF01d} using
more accurate data for the same object at $z=0.247$ and for a more
distant object at $z=0.6847$, and the following limits were
obtained:
\begin{equation}\label{rotCO2}
 \frac{\delta F}{F} = (-2.0\pm 4.4)\times 10^{-6}
\end{equation}
 \begin{equation}\label{rotCO}
 \frac{\delta F}{F} = (-1.6\pm 5.4)\times 10^{-6}
\end{equation}
The object at $z=0.6847$ is associated with the gravitational lens
toward quasar B0218+357 and corresponds to the backward time $\sim
6.5$ Gyr.

\subsection{Enhancement of variation of $\mu$ in
inversion spectrum of ammonia and limit from quasar spectra} \label{secNH3}

Few years ago van Veldhoven et al suggested to use decelerated
molecular beam of ND$_3$ to search for the variation of $\mu$ in
laboratory experiments \cite{ammonia}. Ammonia molecule has a
pyramidal shape and the inversion frequency depends on the
exponentially small tunneling of three hydrogens (or deuteriums)
through the potential barrier. Because of that, it is
very sensitive to any changes of the parameters of the system,
particularly to the reduced mass for this vibrational mode. This fact was used in
\cite{FK1} to place the best limit on the cosmological  variation of $\mu$.

The inversion vibrational mode of ammonia is described by a double
well potential with first two vibrational levels lying below the
barrier. Because of the tunneling, these two levels are split in
inversion doublets. The lower doublet corresponds to the wavelength
$\lambda\approx 1.25$~cm and is used in ammonia masers. Molecular
rotation leads to the centrifugal distortion of the potential curve.
Because of that, the inversion splitting depends on the rotational
angular momentum $J$ and its projection on the molecular symmetry
axis $K$:
 \begin{equation}\label{w_inv}
 \omega_\mathrm{inv}(J,K) = \omega^0_\mathrm{inv}
 - c_1
 \left[J(J+1)-K^2\right] + c_2 K^2 + \cdots \,,
 \end{equation}
where we omitted terms with higher powers of $J$ and $K$.
Numerically, $\omega^0_\mathrm{inv}\approx 23.787$~GHz, $c_1\approx
151.3$~MHz, and $c_2\approx 59.7$~MHz.

In addition to the rotational structure (\ref{w_inv}) the inversion
spectrum includes much smaller hyperfine structure. For the main
nitrogen isotope $^{14}$N, the hyperfine structure is dominated by
the electric quadrupole interaction ($\sim 1$~MHz).
Because of the dipole selection rule $\Delta K=0$ the levels with
$J=K$ are metastable. In astrophysics the lines with $J=K$ are
also narrower and stronger than others, but the hyperfine structure
for spectra with high redshifts is still not resolved.
We obtained the following results for NH$_3$ \cite{FK1} (in atomic units):
\begin{equation}
 \label{dw_inv6}
 \frac{\delta\omega_\mathrm{inv}^0}{\omega_\mathrm{inv}^0}
 \approx 4.46\, \frac{\delta\mu}{\mu}\,.
\end{equation}
 \begin{equation}
 \label{dw_rot5}
 \frac{\delta c_{1,2}}{c_{1,2}}
 = 5.1\frac{\delta\mu}{\mu}\,.
 \end{equation}
For ND$_3$ the inversion frequency is 15 times smaller and this
leads to a higher realative sensitivity of the inversion frequency to
$\mu$:
\begin{equation}
 \label{dw_inv6nd3}
 \frac{\delta\omega_\mathrm{inv}^0}{\omega_\mathrm{inv}^0}
 \approx 5.7\, \frac{\delta\mu}{\mu}\,.
\end{equation}
 \begin{equation}
 \label{dw_rot5nd3}
 \frac{\delta c_{1,2}}{c_{1,2}}
 = 6.2\frac{\delta\mu}{\mu}\,.
 \end{equation}
We see  that the
inversion frequency $\omega_\mathrm{inv}^0$ and the rotational
intervals
$\omega_\mathrm{inv}(J_1,K_1)-\omega_\mathrm{inv}(J_2,K_2)$ have
different dependencies on the constant $\mu$. In principle, this
allows one to study time-variation of $\mu$ by comparing different
intervals in the inversion spectrum of ammonia. For example, if we
compare the rotational interval to the inversion frequency, then
Eqs. (\ref{dw_inv6}) and (\ref{dw_rot5}) give:
\begin{equation}
 \label{red1}
 \frac{\delta\{[\omega_\mathrm{inv}(J_1,K_1)-\omega_\mathrm{inv}(J_2,K_2)]
 /\omega^0_\mathrm{inv}\}}
 {[\omega_\mathrm{inv}(J_1,K_1)-\omega_\mathrm{inv}(J_2,K_2)]/\omega^0_\mathrm{inv}}
 = 0.6 \frac{\delta\mu}{\mu}\,.
\end{equation}
The relative effects are substantially larger if we compare the
inversion transitions with the  transitions between the quadrupole
and magnetic hyperfine components. However, in practice this method
will not work because of the smallness of the hyperfine structure
compared to typical line widths in astrophysics.

We  compared the inversion spectrum of NH$_3$ with
rotational spectra of other molecules, where
\begin{equation}
 \label{red2}
 \frac{\delta\omega_\mathrm{rot}}{\omega_\mathrm{rot}}
 = \frac{\delta\mu}{\mu}\,.
\end{equation}
High precision data on the redshifts of NH$_3$ inversion lines exist
for already mentioned object B0218+357 at $z\approx 0.6847$
\cite{HJK05}. Comparing them with the redshifts of rotational lines
of CO, HCO$^+$, and HCN molecules from Ref.~\cite{CW97} one can get
the following  limit:
\begin{equation} \label{nh3final}
 \frac{\delta\mu}{\mu}=\frac{\delta X_e}{X_e}=(-0.6 \pm 1.9)\times 10^{-6}.
\end{equation}
Taking into account that the redshift $z\approx 0.68$ for the object
B0218+357 corresponds to the backward time about 6.5 Gyr and assuming linear
time dependence , this limit
translates into the most stringent present limit 
for the variation rate $\dot\mu/\mu$ and $X_e$ 
\cite{FK1}:
 \begin{equation}\label{best_mu_dot}
 \dot{\mu}/\mu=\dot{X_e}/X_e=(1 \pm 3) \times 10^{-16}\mathrm{~yr}^{-1}\,.
 \end{equation}
A combination of this result and the atomic clock results
\cite{clock_1,tedesco} gives the best limit on variation of $\alpha$:
 \begin{equation}\label{best_alpha_dot}
 \dot{\alpha}/\alpha=(-0.8 \pm 0.8) \times 10^{-16}\mathrm{~yr}^{-1}\,.
 \end{equation}

\section{Proposals of enhanced effects in diatomic molecules}\label{diatomics}

In transitions between very close narrow levels of different
nature in diatomic molecules the relative effects of the variation
may be enhanced by several orders of magnitude. Such levels may occur due to
cancelation between either hyperfine and rotational structures
\cite{mol}, or between the fine and vibrational structures of the
electronic ground state \cite{FK2}. The intervals between the levels
are conveniently located in microwave frequency range and the level
widths are very small, typically $\sim 10^{-2}$~Hz.

\subsection{Molecules with cancelation between hyperfine
structure and rotational intervals} \label{hfs-rot}

Consider diatomic molecules with unpaired electron and ground state
$^2\Sigma$. It can be, for example, LaS, LaO, LuS, LuO, YbF,
etc.~\cite{HH79}. Hyperfine interval $\Delta_\mathrm{hfs}$ is
proportional to $\alpha^2 Z F_\mathrm{rel}(\alpha Z) \mu
g_\mathrm{nuc}$, where $F_\mathrm{rel}$ is additional relativistic
(Casimir) factor.
 Rotational interval
$\Delta_\mathrm{rot} \sim \mu$ is roughly independent on $\alpha$.
If we find molecule with $\Delta_\mathrm{hfs} \approx
\Delta_\mathrm{rot}$ the splitting $\omega$ between hyperfine and
rotational levels will depend on the following combination
\begin{equation}
\label{hfs-rot1}
 \omega \sim  \left[\alpha^2 F_\mathrm{rel}(\alpha Z)\, g_\mathrm{nuc}
 - \mathrm{const}\right]\, .
\end{equation}
Relative variation is then given by
\begin{equation}
\label{hfs-rot2}
 \frac{\delta\omega}{\omega}
 \approx \frac{\Delta_\mathrm{hfs}}{\omega}
 \left[\left(2+K\right)\frac{\delta\alpha}{\alpha} + \frac{\delta
 g_\mathrm{nuc}}{g_\mathrm{nuc}}\right]\,,
\end{equation}
where factor $K$ comes from variation of $F_\mathrm{rel}(\alpha Z)$,
and for $Z \sim 50$, $K\approx 1$.
 Using data from
\cite{HH79} one can find that $\omega = (0.002\pm 0.01)$~cm$^{-1}$
for ${}^{139}$La${}^{32}$S \cite{mol}. Note that for $\omega =
0.002$~cm$^{-1}$ the relative frequency shift is:
\begin{equation}
\label{hfs-rot3}
 \frac{\delta\omega}{\omega}
 \approx 600\,\frac{\delta\alpha}{\alpha}\,.
\end{equation}


\subsection{Molecules with cancelation between fine
structure and vibrational intervals} \label{fs-vib}

The fine structure interval $\omega_f$ rapidly grows with nuclear
charge Z:
\begin{equation}
\label{of}
 \omega_f \sim Z^2 \alpha^2\, ,
\end{equation}
The vibration energy quantum decreases with the
atomic mass:
\begin{equation}
 \label{ov}
\omega_\mathrm{vib} \sim M_r^{-1/2} \mu^{1/2}\, ,
\end{equation}
where the reduced mass for the molecular vibration is $M_r m_p$.
Therefore, we obtain equation $Z=Z(M_r,v)$ for the lines on the
plane $Z,M_r$, where we can expect approximate cancelation between
the fine structure and vibrational intervals:
\begin{equation}
 \label{o}
 \omega=\omega_f - v\,  \omega_\mathrm{vib} \approx 0 \,,
 \quad v=1,2,...
\end{equation}
Using Eqs.~(\ref{of}--\ref{o}) it is easy to find dependence of the
transition frequency on the fundamental constants:
\begin{equation}
 \label{do}
 \frac{\delta\omega}{\omega}=
 \frac{1}{\omega}\left(2 \omega_f \frac{\delta\alpha}{\alpha}+
 \frac{v}{2} \omega_\mathrm{vib} \frac{\delta\mu}{\mu}\right)
 \approx K \left(2 \frac{\delta\alpha}{\alpha}+
\frac{1}{2} \frac{\delta\mu}{\mu}\right),
\end{equation}
where the enhancement factor $K= \frac{\omega_f}{\omega}$
determines the relative frequency shift for the given change of
fundamental constants. Large values of factor $K$ hint at
potentially favorable cases for making experiment, because it is
usually preferable to have larger relative shifts. However, there is
no strict rule that larger $K$ is always better. In some cases, such
as very close levels, this factor may become irrelevant. Thus, it is
also important to consider the absolute values of the shifts and
compare them to the linewidths of the corresponding transitions.

Because the number of molecules is finite we can not have $\omega=0$
exactly. However, a large number of molecules have $\omega/\omega_f
\ll 1$ and $|K| \gg 1$. Moreover, an additional ``fine tuning'' may
be achieved by selection of isotopes and rotational,
$\Omega$-doublet, and hyperfine components. Therefore, we have two
large manifolds, the first one is build on the electron fine
structure excited state and the second one is build on the
vibrational excited state. If these manifolds overlap one may select
two or more transitions with different signs of $\omega$. In this
case expected sign of the $|\omega|$-variation must be different
(since the variation $\delta \omega$ has the same sign) and one can
eliminate some systematic effects. Such control of systematic
effects was used in \cite{budker,budker1} for transitions between
close levels in two dysprosium isotopes. The sign of energy
difference between two levels belonging to different electron
configurations was different in $^{163}$Dy and $^{162}$Dy.

Among the interesting molecules  where the ground state is split in two fine
structure levels and (\ref{o}) is approximately fulfilled, there are
Cl$_2^+$ (enhancement $K=1600$),  SiBr ($K=360$), CuS ($K=24$) and 
 IrC ($K=160$). The list of molecules  is not complete because of the
lack of data in \cite{HH79}. The
molecules Cl$_2^+$ and SiBr are particularly interesting. For both
of them the frequency $\omega$ defined by (\ref{o}) is of the order
of 1~cm$^{-1}$ and comparable to the rotational constant $B$. That
means that $\omega$ can be reduced further by the proper choice of
isotopes, rotational quantum number $J$ and hyperfine components, so
we can expect $K \sim 10^3-10^5$. New dedicated measurements are
needed to determined exact values of the transition frequencies and
find the best transitions. However, it is easy to  find necessary
accuracy of the frequency shift measurements. According to (\ref{do})
the expected frequency shift is
\begin{equation}
\label{do1}
 \delta\omega=2 \omega_f \left(\frac{\delta\alpha}{\alpha}+
 \frac{1}{4}\frac{\delta\mu}{\mu}\right)
\end{equation}
Assuming $\delta \alpha / \alpha \sim 10^{-15}$ and $\omega_f\sim
500$~cm$^{-1}$, we obtain $\delta\omega \sim 10^{-12}$ cm$^{-1}\sim
3 \times 10^{-2}$ Hz  (in order to obtain similar sensitivity
comparing hyperfine transition frequencies for Cs and  Rb one has to
measure the shift $\sim 10^{-5}$ Hz). This shift is larger than the
 natural width $\sim 10^{-2}$ Hz.

\subsection{Molecular ion {H\lowercase{f}F}$^+$}

 The ion HfF$^+$ and other
similar ions are considered by Cornell's group in JILA for the
experiment to search for the electric dipole moment (EDM) of the
electron. Recent calculation by
\cite{PMI06} suggests that the ground state of this ion is
$^1\Sigma^+$ and the first excited state $^3\Delta_1$ lies only
1633~cm$^{-1}$ higher. Calculated vibrational frequencies for these
two states are 790 and 746~cm$^{-1}$ respectively. For these
parameters the vibrational level $v=3$ of the ground state is only
10~cm$^{-1}$ apart from the $v=1$ level of the state $^3\Delta_1$.
Thus, instead of (\ref{o}) we now have:
\begin{equation}
 \label{hff1}
 \omega=\omega_\mathrm{el} + \frac32 \omega_\mathrm{vib}^{(1)}
 - \frac72\omega_\mathrm{vib}^{(0)}\approx 0\,,
\end{equation}
where superscripts 0 and 1 correspond to the ground and excited
electronic states. Electronic transition $\omega_\mathrm{el}$ is not
a fine structure transition and (\ref{of}) is not applicable.
Instead,  we can write:
\begin{equation}
 \label{hff2}
 \omega_\mathrm{el}=\omega_\mathrm{el,0} + q x\,,
 \quad x=\alpha^2/\alpha_0^2-1\,.
\end{equation}
Our estimate is \cite{FK2}
\begin{equation}
 \label{hff3}
 \frac{\delta\omega}{\omega}
 \approx
 \left(\frac{2q}{\omega} \frac{\delta\alpha}{\alpha}+
 \frac{\omega_\mathrm{el}}{2\omega} \frac{\delta\mu}{\mu}\right)
 \approx \left(2000 \frac{\delta\alpha}{\alpha}+
 80 \frac{\delta\mu}{\mu}\right),
\end{equation}
\begin{equation}
 \label{hff4}
 \delta\omega
 \approx 20000~\mathrm{cm}^{-1}(\delta\alpha/\alpha+0.04 \delta\mu/\mu)\,.
\end{equation}
Assuming $\delta \alpha / \alpha \sim 10^{-15}$ we obtain
$\delta\omega \sim$ 0.6 Hz. The natural width is about 2 Hz.

We also present the result for transition between close levels
in Cs$_2$ molecule suggested in \cite{DeMille,DeMille1}. Our estimate is
\cite{kozlovmol}:
\begin{equation}\label{Cs2c}
 \delta\omega \approx (-240\frac{\delta\alpha}{\alpha}
 -1600\frac{\delta\mu}{\mu}) cm^{-1}\,,
\end{equation}

\section{Changing physics near massive bodies}
In this section I follow Ref. \cite{FS2007}.

The reason gravity is so important at large scales is that
its effect is additive. The same should be true for
 massless (or very light) scalars: its effect near large body is proportional to the
 number of particles in it.
  
For not-too-relativistic objects, like the usual stars or planets,
 both their total  
mass $M$ and the total scalar charge $Q$ are simply proportional to
the number of nucleons in them, and thus the scalar field is simply
proportional to the gravitational potential
\begin{equation} \label{kappa}
 \phi-\phi_0=\kappa (GM/rc^2) \,. 
\end{equation}
Therefore, we expect that the fundamental constants would also depend on
the
position via the gravitational potential at the the measurement point.

Gravitational potential on Earth is changing due to ellipticity of its
orbit:
the corresponding variation of the Sun graviational potential is
 $\delta (GM/rc^2)=3.3 \cdot 10^{-10}$. The
accuracy of atomic clocks in laboratory conditions is about
$10^{-16}$. As an example we consider recent
work \cite{clock_1} who obtained the following value for the half-year
variation
of the frequency ratio of two atomic clocks: (i) optical transitions in 
mercury ions $^{199}Hg^+$ and (ii) hyperfine splitting
in $^{133}Cs$ (the frequency standard). The limit obtained is
\begin{equation}
 \delta ln({\omega_{Hg}\over \omega_{Cs}})=(0.7\pm 1.2) \cdot 10^{-15}
\end{equation}  
For  Cs/Hg  frequency ratio of these clocks  the dependence on the fundamental
constants
was evaluated in \cite{tedesco} with the result
\begin{equation}
 \delta ln({\omega_{Hg}\over \omega_{Cs}})=-6 {\delta \alpha \over
\alpha} -0.01{\delta  (m_q/\Lambda_{QCD}) \over (m_q/\Lambda_{QCD})} -{\delta  (m_e/M_p) \over (m_e/M_p)}
\end{equation}
Another work \cite{BW} compare $H$ and $^{133}Cs$ hyperfine transitions.
The amplitude of the half-year variation  found were
\begin{equation}
 |\delta ln(\omega_{H}/\omega_{Cs})| <7 \cdot 10^{-15}
\end{equation}
The sensitivity \cite{tedesco}
\begin{equation}
\delta ln({\omega_{H}\over \omega_{Cs}})=-0.83 {\delta \alpha \over
\alpha} -0.11{\delta  (m_q/\Lambda_{QCD}) \over (m_q/\Lambda_{QCD})}
\end{equation}
There is no sensitivity to $m_e/M_p$ because
they are both hyperfine transitions.

As motivated above, we assume that scalar and gravitational
potentials are proportional to each other,  and thus introduce parameters
$k_i$ as follows
\begin{equation}
 {\delta \alpha \over \alpha } = k_\alpha \delta ({GM\over r c^2})
\end{equation}
\begin{equation}
  {\delta (m_q/\Lambda_{QCD}) \over (m_q/\Lambda_{QCD})} = k_q \delta ({GM\over r c^2})
\end{equation}
\begin{equation}
  {\delta (m_e/\Lambda_{QCD}) \over (m_e/\Lambda_{QCD}) } = {\delta (m_e/M_p) \over (m_e/M_p) } =k_e \delta ({GM\over r c^2})
\end{equation}
where in the r.h.s. stands half-year variation of Sun's gravitational potential
on Earth.

In such terms, the results of  Cs/Hg  frequency ratio measurement
\cite{clock_1}  can be rewritten as
\begin{equation}
 k_\alpha +0.17 k_e= (-3.5\pm 6) \cdot 10^{-7}
\end{equation}
 The results of  Cs/H  frequency ratio measurement \cite{BW} can be presented as 
\begin{equation}
| k_\alpha +  0.13 k_q | <2.5 \cdot 10^{-5}
\end{equation}
Finally, the result of recent measurement  \cite{Ashby} of
  Cs/H  frequency ratio can be presented as 
\begin{equation}
 k_\alpha +0.13 k_q= (-1\pm 17) \cdot 10^{-7}
\end{equation}
The sensitivity coefficients for other  clocks have been discussed above.

Two new results have been obtained recently. From transition between
close levels in Dy we obtained \cite{budkerG}
\begin{equation}
 k_\alpha = (-8.7\pm 6.6) \cdot 10^{-6}
\end{equation}
From optical  Sr/hyperfine Cs comparison we obtained \cite{YeG}
\begin{equation}
 k_\alpha+0.36 k_e = (1.8\pm 3.2) \cdot 10^{-6}
\end{equation}
Combination of the data gives \cite{YeG}
\begin{equation}
 k_\alpha = (-2.3\pm 3.1) \cdot 10^{-6}
\end{equation}
\begin{equation}
 k_e = (1.1\pm 1.7) \cdot 10^{-5}
\end{equation}
\begin{equation}
 k_\alpha = (1.7\pm 2.7) \cdot 10^{-5}
\end{equation}

\section{Acknowledgments}
The author is grateful to E. Shuryak and M. Kozlov
 for valuable contributions to this review.
 This work is supported by the Australian
 Research Council.

%

%
%

\end{document}